\documentclass[a4paper, portrait]{spie}  

\usepackage[dvips]{graphicx} 
\usepackage{epsfig}
\usepackage{subfigure}

\title{A study on the use of the PACS bolometer arrays for submillimeter ground-based telescopes}

\author{Vincent Reveret\supit{a,b},  Louis R. Rodriguez\supit{b}, Philippe Andr\'e\supit{b}, Benoit Horeau\supit{b}, Jean Le Pennec\supit{b}, Patrick Agn\`ese\supit{c}
\skiplinehalf
\supit{a}European Southern Observatory, Casilla 19001, Santiago 19, Chile \\
\supit{b}CEA/DSM/DAPNIA Service dÕAstrophysique, Saclay, 91191 Gif sur Yvette Cedex, France  \\
\supit{c}LETI/CEA Grenoble, 17 avenue des Martyrs, 38054 Grenoble Cedex 9, France
}

\authorinfo{Contact author at :  vreveret@eso.org}
 
\begin{document} 
  \maketitle 

\begin{abstract}
A new kind of bolometric architecture has been successfully developed for the PACS photometer onboard the Herschel submillimeter observatory. These new generation CCD-like arrays are buttable and enable the conception of large fully sampled focal planes. We present a feasibility study of the adaptation of these bolometer arrays to ground-based submillimeter telescopes. We have developed an electro-thermal numerical model to simulate the performances of the bolometers under specific ground-based conditions (different wavelengths and background powers for example). This simulation permits to determine the optimal parameters for each condition and shows that the bolometers can be background limited in each transmission window between 200 and 450 microns. We  also present a new optical system that enables to have a maximum absorption of the bolometer in each atmospheric windows. The description of this system and measurements are showed.  
\end{abstract}

\keywords{Large bolometer array, ground-based telescopes, submillimeter astronomy, simulations, submillimetric absorption, thin dielectric layers.}

\section{INTRODUCTION - CEA Bolometers.}
\label{sect:intro}  

CEA has now a strong experience in the domain of large bolometer arrays for submillimeter astronomy. In 1997, LETI/LIR and DAPNIA/SAp have started to develop new technology bolometers  to meet the requirements of the HERSCHEL Space Observatory (see G.L. Pilbratt, this conference vol. 6265) : how to build a wild-field submillimetric camera with  fast mapping capability and a very good sensitivity  in a relatively high background power environment? 

The original approach was to use different known technologies, many of them developed during the ISOCAM project at CEA (Ref. [\citenum{Cesarsky96}]), and put them together to build a new kind of bolometer array. Such technologies include ionic implantation (for resistive thermometers), silicon micro-machining, flipchip technology using indium bumps for bonding the different elements and low temperature CMOS amplification. It was also decided not to use Winston cones, but rather a "resonant metallic absorption" system.

That led to the conception of a non traditionnal bolometric detection system, more similar to the CCD concepts used in other domains of astrophysics. 256 pixels compose the basic array, each pixel having a 0.5F$\lambda$ field of view to fully sample the image in the focal plane (see F. Simoens et al., this volume).  Pixels contain  two semiconducting thermometers (silicon made with phosphorus implantation and boron compensation) which work at 300 mK in the "hopping conduction"  regime. The submillimeter light absorption is obtained with a quarter wave cavity. The bottom of the cavity is a silicon layer coated with a thin gold layer that acts as a reflector (more details in section 3). Indium bumps support the upper absorbing layer made of silicon  which is metal coated  (TiN, superconductive at 300 mK but absorbing for submillimeter frequencies). This layer has a grid-like geometry  to reduce the heat capacity of the bolometer.

\begin{figure}
   \begin{center}	
      	\includegraphics[height=6.75cm]{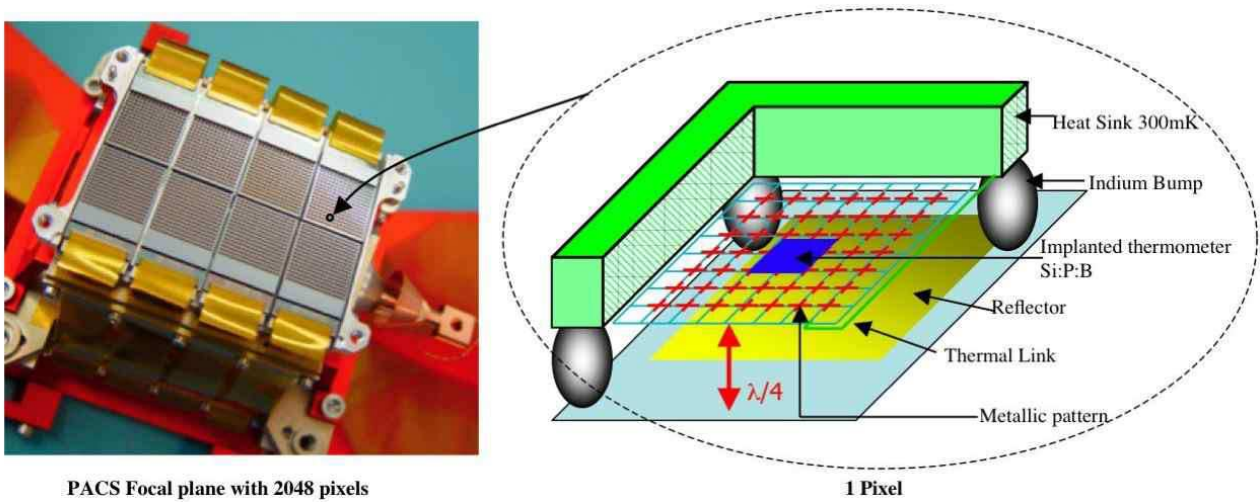}	
   \end{center}
   \caption[schemapix] 
   { \label{schemapix} 
The picture on the left shows a pre-version of the 100 $\mu$m focal plane of the HERSCHEL PACS photometer. It is made of 8 subarrays each containing 256 bolometers. The sketch on the right side describes one of these pixels. A first electronic stage (impedance reduction) is located below the quarter wave cavity reflector, at 300 mK. CMOS transistors at 2K located a few cm below the pixels perform the multiplexing.}
   \end{figure}

The signal is read out at the middle point of the resistor bridge and is processed by CMOS transistors. Because of the intrinsic noise of CMOS followers, the impedances of the bolometer resistors have to be  very high (a few G$\Omega$) in order to provide a very high response. A cold (2K) CMOS  multiplexing system (16 to 1) is located below the detection stage. These bolometer arrays can be butted end to end to form very large focal planes (see figure \ref{schemapix}).

CEA has built two bolometric focal planes for the PACS photometer on HERSCHEL (see N. Billot et al., vol. 6265 this conference). Flight models are being tested and calibrated. The detectors fulfill all the requirements in particular in terms of sensitivity (BLIP\footnote{BLIP : Background Limited Performances} conditions), bandwidth (electrical and thermal) and 1/$f$ noise suppression.

Since 2000, we have started to work on the ways to adapt these bolometer arrays to ground-based submillimeter telescopes. Indeed, this now well-mature  bolometer technology seems very well suited for ground-based conditions, like high and variable background power on the focal plane or large field of view coverage. Recent developments are presented in the following sections.

\section{Electro-Thermal Model} 

\subsection{Description} 
\label{sect:title}

In order to estimate the best performances for our bolometers, we have developed a numerical and dynamical model  which permits to obtain optimum values of different parameters (such as bias, or thermal conductance for example) in different observing conditions. The model is based on the general differential equation that manages thermal exchanges in the bolometer (see Ref. [\citenum{Richards94}]) :
 \begin{equation}
P_{ph}+P_{J}=C\frac{dT_b}{dt}+ \overline{G_{th}}(T_b-T_0)
\end{equation}

with $P_{ph}=$ total photonic power incident on one pixel, $P_{J}=$ electrical power coming from the bias of the resistive element in the bolometer, $C=$ total heat capacity of the pixel, $T_b=$ bolometer temperature, $T_0=$ heat sink temperature, constant at 300 mK, $\overline{G_{th}}=$ average value of the thermal conductance between $T_b$ and $T_0$, and $t=$ time.

\begin{figure}
   \begin{center}
   \begin{tabular}{c}	
   	\subfigure[]{\includegraphics[bb=57 285 501 555, height=6cm]{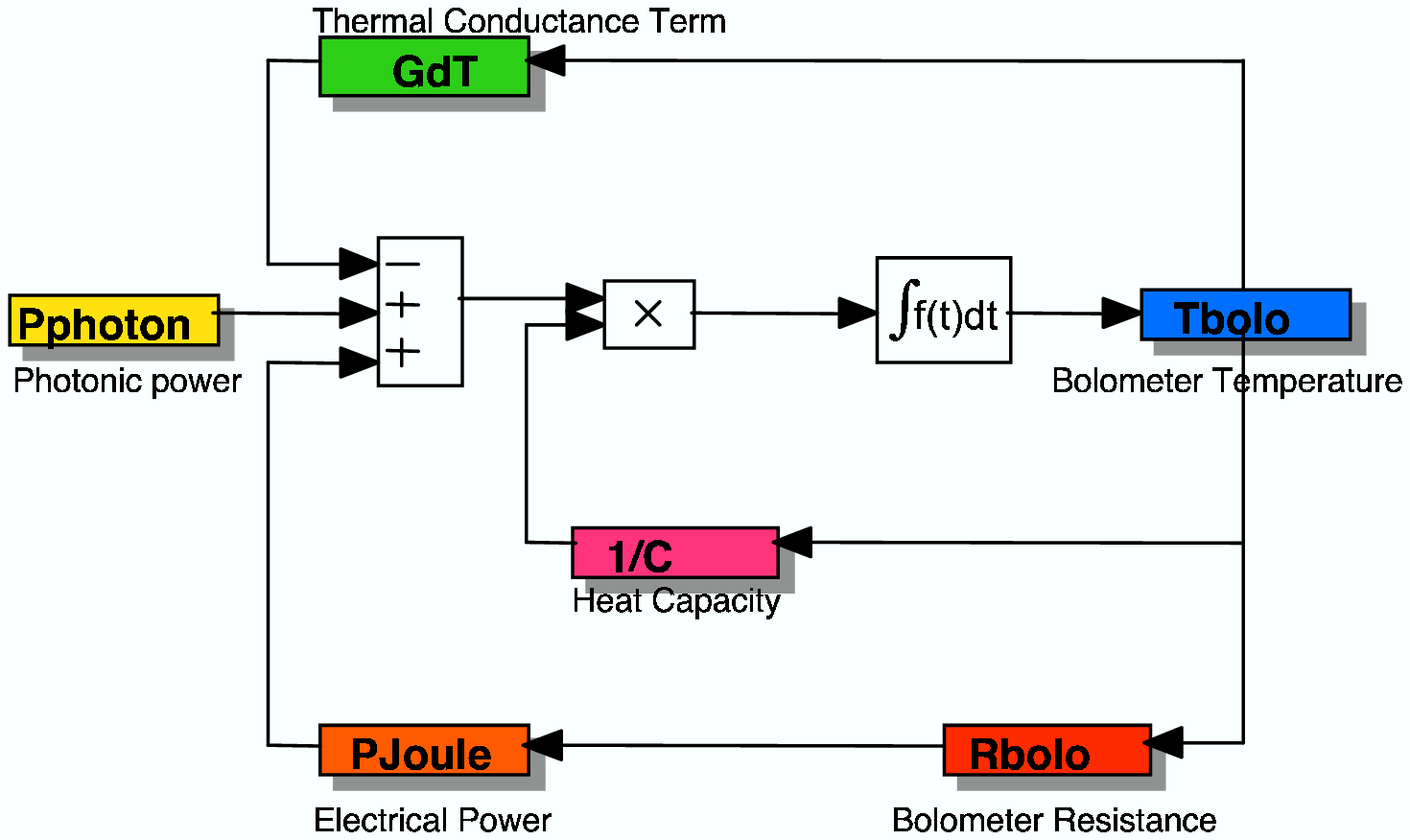}}	
	\subfigure[]{\includegraphics[bb=479 56 769 322, height=6cm]{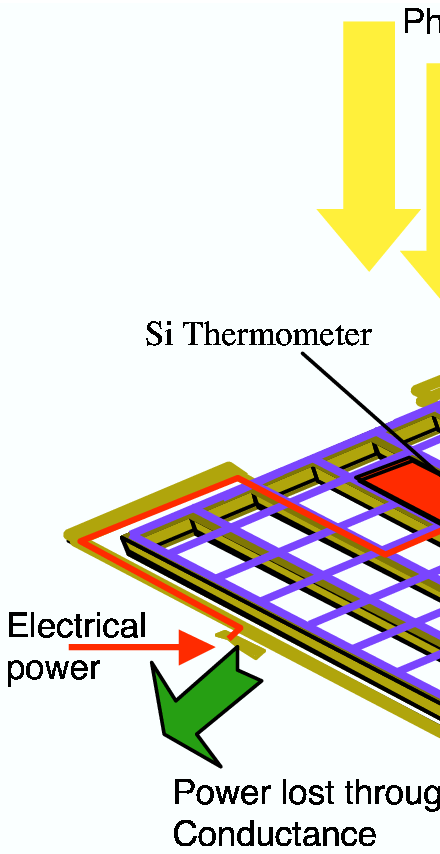}}  
   \end{tabular}
   \end{center}
   \caption[example] 
   { \label{model} 
(a) Principle of the electro-thermal dynamical model used to simulate the bolometer behaviour. It is based on the power balance inside the bolometer and includes specific parameters such as  heat capacity,  thermal conductance and empiric relations for the resistor behaviour. (b) Schematic representation of the different powers involved in the bolometric detection.}
   \end{figure}

Solving this equation permits to get the temperature of the bolometer as a function of time (see figure \ref{model}). All the other parameters depend on this temperature  :

\begin{itemize}

\item Bolometer impedance (empirically adapted from Efros Law, ref.[\citenum{Buzzi99}])  :

\begin{equation}
R=R_0\exp\left(\sqrt{\frac{T_0}{T}}\right)\exp\left(-\frac{qL_{(T)}E}{kT}\right)
\label{eqefros}
\end{equation}

where $R_0$ and $T_0$ are constants caracterising the type of resistor we are using (Silicon type with implanted Phosphorus and Boron compensation), $L_{(T)}$ is the hopping length and $E$ is the electric field inside the resistor.

\item Thermal conductance :

\begin{equation}
G(T)=\alpha_1T+\beta_1T^3
\label{eqconductth}
\end{equation}

where the $T$ term characterizes the heat conduction in metals (from Drude theory) and the $T^3$ term corresponds to a dielectric conduction (Casimir model).

\item Heat capacity :

\begin{equation}
C(T)=\alpha_2T+\beta_2T^3+\gamma_1\exp\left(\frac{cste}{T}\right)
\end{equation}

where the $T$ term corresponds to the doped silicon resistor, the $T^3$ term corresponds to the dielectric (silicon grid) and the exponential term is a characteristics of the absorbing metal (TiN).

\end{itemize}

All the constant terms and empiric laws for these parameters have been determined experimentally. The model itself has been created using MATLAB\footnote{MATLAB is a mathematical software from \textit{The MathWorks} Company} and its extension SIMULINK. We use this simulation to determine what would be the best performances  of our bolometers in ground-based conditions.

\subsection{Atmospheric Bands} 

\begin{figure}
   \begin{center}
   \begin{tabular}{c}		
   \includegraphics[height=11cm, angle=90]{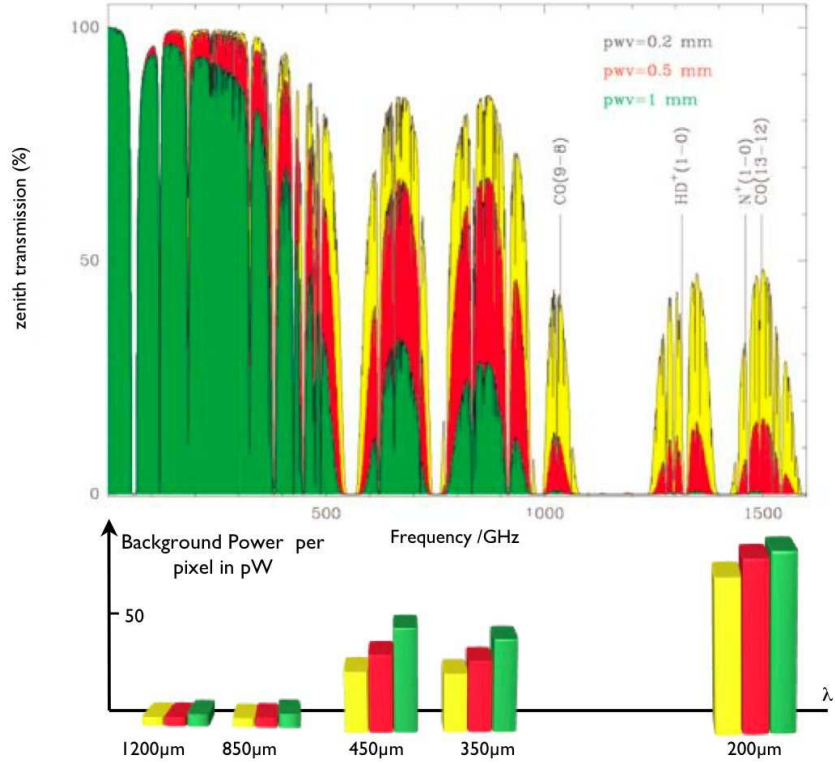}	
   \end{tabular}
   \end{center}
   \caption[example] 
   { \label{atmo} 
The different  atmospheric transmission windows in the submillimeter regime on the Chajnantor site in the Atacama desert. The  colors correspond to  different values of precipitable water vapour contents (adapted from [\citenum{Nyman02}]). The bottom part shows the estimated background power falling on one pixel, mainly coming from the atmosphere and the telescope.}
   \end{figure}

Only a few sites on Earth are suitable for submillimeter astronomy between 200 and 865 $\mu$m. The Chajnantor plateau in the Atacama desert for example provides very good conditions, even for the shortest wavelengths. Figure  \ref{atmo} shows the typical atmospheric transmission curves on that site for different amounts of precipitable water vapour. We have investigated the behaviour of our bolometers in the main following atmospheric windows: 200 $\mu$m, 350 $\mu$m, 450 $\mu$m, 865 $\mu$m and 1.2 mm considering a 12m diameter telescope on a site like Chajnantor.

For each of these bands, we have estimated the background power incoming on one pixel. The total background power, $P_{pix}$, is calculated by :

\begin{equation}
P_{pix}=E  \eta_{pix}\eta_{abs}\left(P_{opt}+P_{tel}+P_{sky}\right)
\label{pbackgnd}
\end{equation}

where $E=$ throughput of the system, $\eta_{pix}=$ geometric filling factor of a pixel, $\eta_{abs}=$ absorption coefficient of the pixel, $P_{opt}=$  power emitted by the cold internal optics of the camera, $P_{tel}=$ power emitted by the telescope and $P_{sky}=$ power coming from the atmosphere (depends on opacity). Typical values of background power are shown in figure \ref{atmo} and table 1.

\subsection{Results of Simulations} 

One of the possible ways to use the numerical model is to determine the optimum voltage of the bolometer which gives the best sensitivity. To estimate it, we add to the constant optical background a very small low frequency power component and see how the output parameters evolve when increasing the voltage bias. Amongst these output parameters, we use the NEP (Noise Equivalent Power) as a figure of merit for the bolometer sensitivity, and also the electro-optical response, in Volt per Watt.

\begin{figure}
   \begin{center}
   \begin{tabular}{c}	
   	\subfigure[]{\includegraphics[bb=71 211 535 588, height=6.5cm]{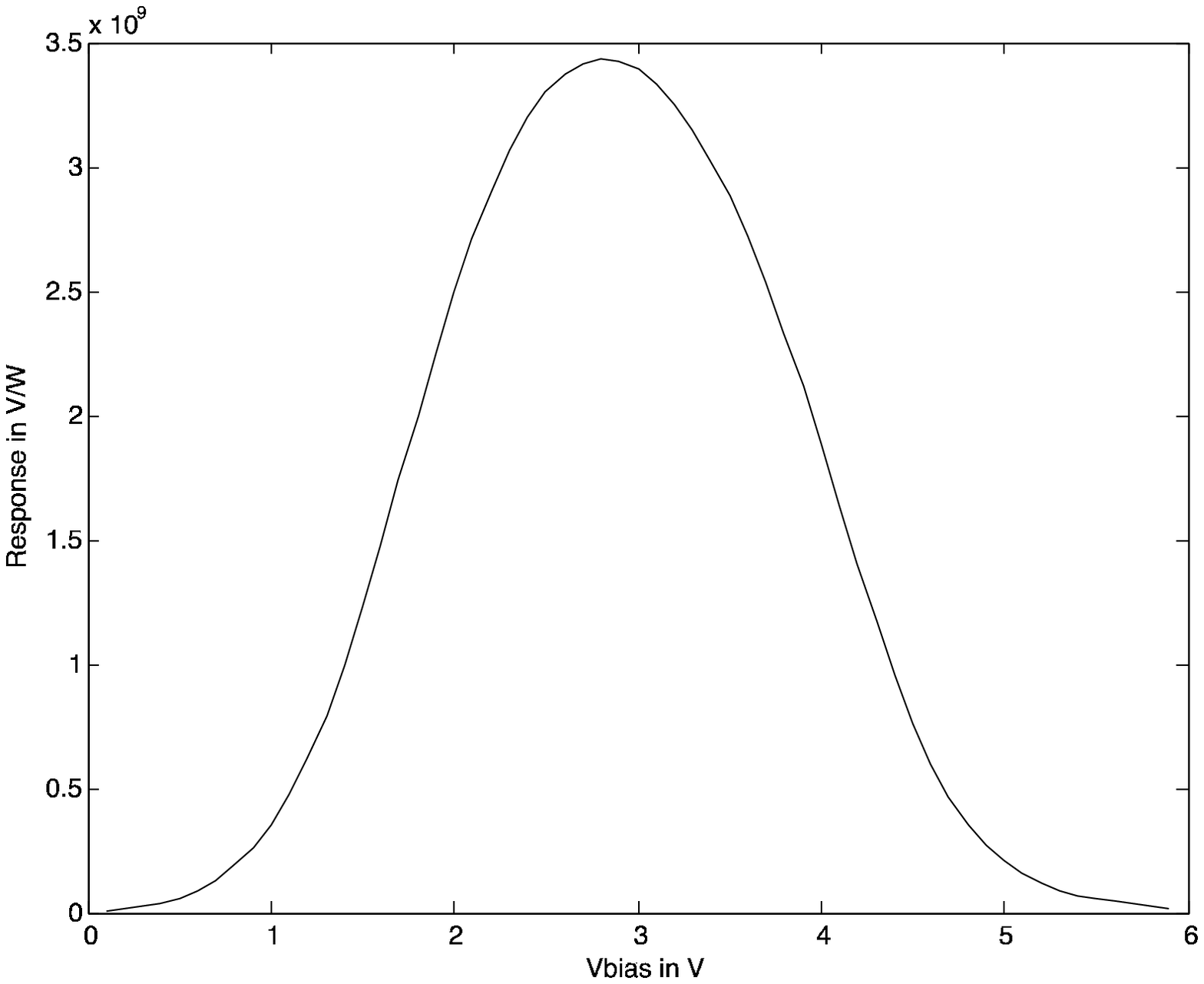}}
	\subfigure[]{\includegraphics[bb=60 210 534 586, height=6.5cm]{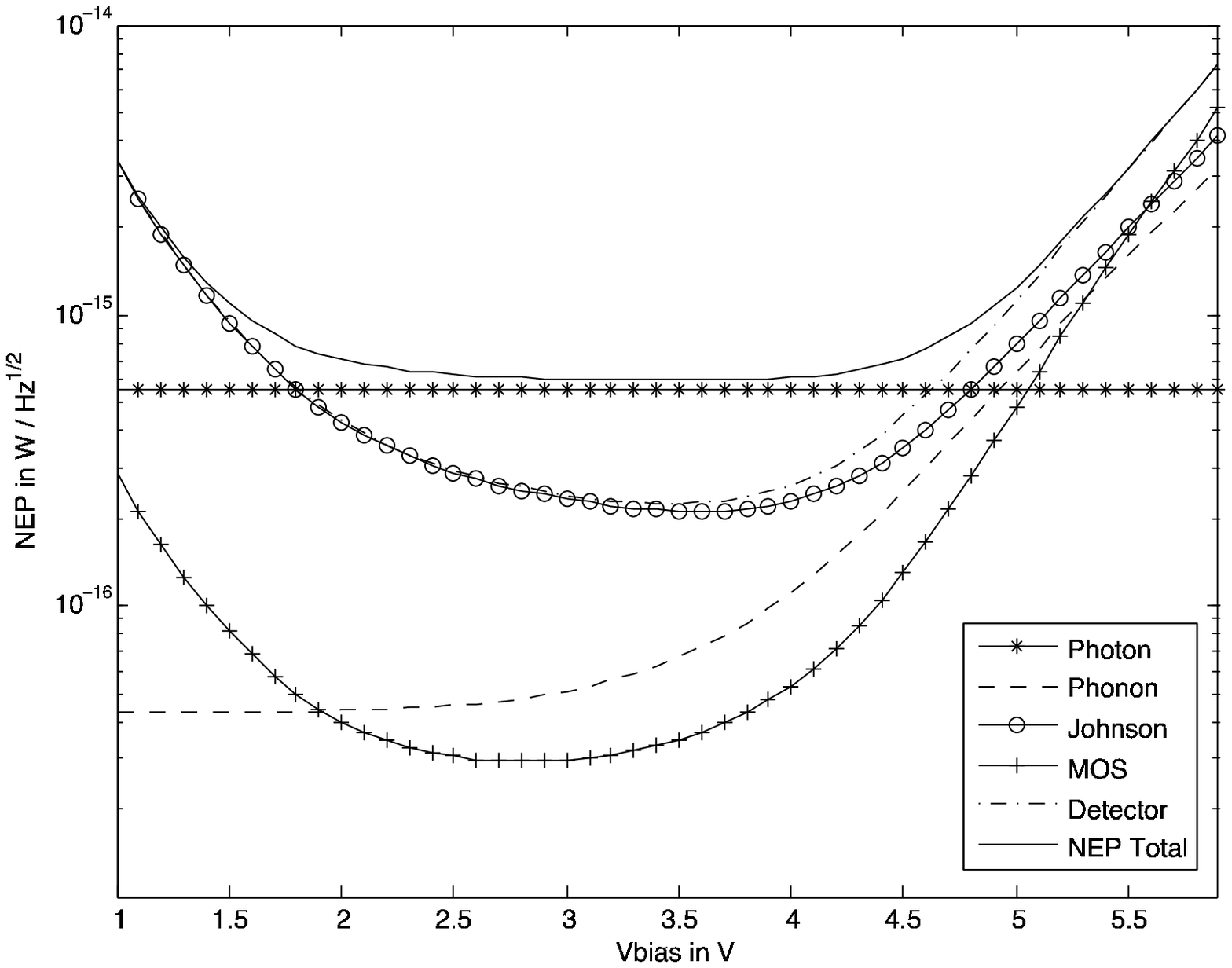}}
   \end{tabular}
   \end{center}
   \caption[example] 
   { \label{nep200} 
(a) Simulated electro-optical response of the bolometer vs resistor bridge voltage, in the case of a power background corresponding to the 200 $\mu$m atmospheric window.  (b) Simulated NEP as function of the bias in the same background conditions.  The optimum bias corresponding to the minimum detector NEP is in that case 3.5 V.}
   \end{figure}

The figure \ref{nep200}b shows the typical NEP curves as functions of the bolometer voltage, in the case of a use in the 200 $\mu$m window. We can see also that the electro-optical response of the bolometer (figure \ref{nep200}a) reaches a maximum for almost the same value of bias. This curves shapes can be explained by the following : 

\begin{itemize}
\item (a) before the optimum bias point, the electro-optical response increases with the voltage bias, as for a normal resistor behaviour. The bolometer temperature increases slightly, which in return decreases the resistance of the bolometer. This effect decreases the noise associated to the bolometer resistor (Johnson noise).
\item (b) after the optimum bias point, even if the thermal conductance increases, it is not sufficient enough to evacuate all the accumulated heat (the heat capacity also depends on the temperature). There is a thermal runaway due to the rapid increase of the bolometer temperature which produces a rapide degradation of the performances.

\end{itemize}

 \begin{figure}
   \begin{center}
   \begin{tabular}{c} 
   \includegraphics[bb=166 78 683 376, height=6cm]{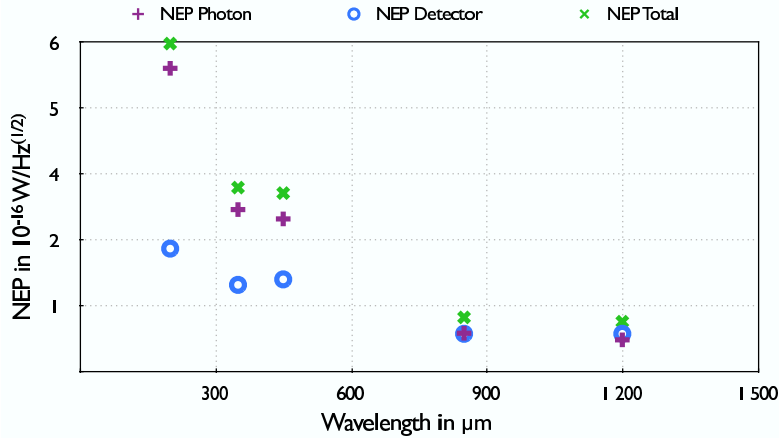}
   \end{tabular}
   \end{center}
   \caption[example] 
   { \label{neps} 
Estimated values of the minimum NEP for the 5 atmospheric windows considered earlier.}
   \end{figure}

For each atmospheric window, we are therefore able to determine the optimum bias that leads to the best NEP. Note that, we also tried to optimize the value of the thermal conductance and found that the best performances are obtained for a small change of the value compared to the one used for PACS  ($\overline{G_{th}}$ increased by a factor of 1.2).

The figure \ref{neps} shows the different NEP in each atmospheric band. It shows that for the three shorter wavelengths  bands (200, 350 and 450 $\mu$m) the performances are clearly limited by the background. It means that  we could use a single bolometer to get BLIP conditions in these three bands.   The 865 $\mu$m and 1.2 mm bands would need another type of bolometer, with a thermal conductance more adapted to the lower background conditions. 

Table 1 gives the estimated NEP and NEFD for each considered atmospheric band. The NEFD (Noise Equivalent Flux Density) is an estimation of the instrument sensitivity during an astronomical observation. It is equivalent to the flux of an object that is detected with a Signal / Noise = 1 in 1s of integration and is related to  the telescope diameter, the optics transmission and the atmospheric conditions.  In the table, it is given for the usual 5$\sigma$, 1 hour integration conditions.
At 200 $\mu$m, this (point) source detection limit is about 28 mJy. For comparaison, the PACS sensitivity will be about 3 mJy. The big advantage of such a ground-based instrument over a space mission is of course its angular resolution, which in a case of a 12m diameter telescope would be $\sim$4 arcsec at 200 $\mu$m ($\sim$13 arcsec for PACS). Because of the existing possibility to join the arrays, it would be possible to build the same kind of focal plane as the ones found in PACS,  (32$\times$64 pixels), to cover  relatively large fields of view in all the bands.

\begin{table}[htbp]
  \begin{center}
\begin{tabular}{|*{6}{c|}}
   \hline
   Wavelength in $\mu$m & 200 & 350 & 450 & 850 & 1200 \\
   \hline
   Average Window Transmission & 50\% & 65\% & 65\% & 85\% & 90\% \\
   \hline
   Typical Background per pixel in pW & 70 & 35 & 40 & 5 & 5 \\
   \hline
    Beam FWHM in arcsec & 4 & 7.5 & 9.5 & 18 & 25 \\
   \hline
    Field of View (for a 32$\times$64 array) & 0.9' $\times$ 1.8' & 1.6' $\times$ 3.2' & 2.1' $\times$ 4.2' & 3.9' $\times$ 7.8' & 5.5' $\times$ 11' \\
   \hline
   NEP Photon in W/Hz$^{1/2}$ & 5.5 $\times10^{-16}$ & 2.9 $\times10^{-16}$ & 2.8 $\times10^{-16}$ & 0.7 $\times10^{-16}$ & 0.6 $\times10^{-16}$ \\
   \hline
   NEP Detector in W/Hz$^{1/2}$ & 2.2 $\times10^{-16}$ & 1.6 $\times10^{-16}$ & 1.7 $\times10^{-16}$ & 0.7 $\times10^{-16}$ & 0.7 $\times10^{-16}$ \\
   \hline
   NEP Total in W/Hz$^{1/2}$ & 6.0 $\times10^{-16}$ & 3.3 $\times10^{-16}$ & 3.3 $\times10^{-16}$ & 1.2 $\times10^{-16}$ & 1.1 $\times10^{-16}$ \\
   \hline
   NEFD in mJy (5$\sigma$, 1h) & 28 & 9 & 5 & 3 & 1.5 \\ 
   \hline
  \end{tabular}
\caption[]{Estimated performances of the bolometers in the different atmospheric windows.} 
\label{tableInstrus}
  \end{center}
\end{table}

\section{Absorption System} \label{sect:sections}
\subsection{Problem and Solution} 

\begin{figure}
   \begin{center}
   \begin{tabular}{c}	
   \includegraphics[bb=160 34 435 813, width=6cm, angle=90]{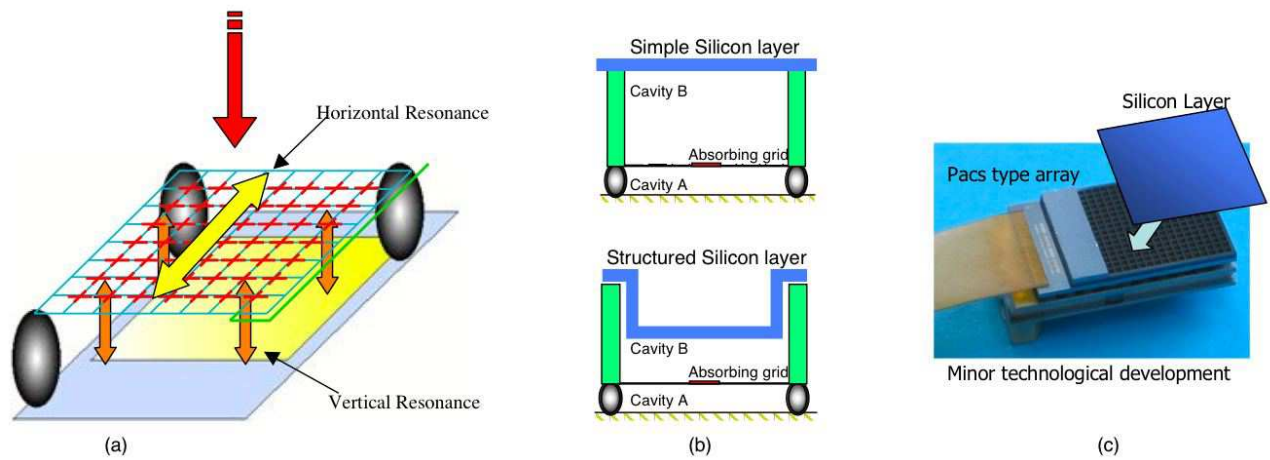}	
   \end{tabular}
   \end{center}
   \caption[example] 
   { \label{absprinciple} 
(a) Schematic representation of the "double resonance" absorption system of one pixel. (b) Representations of two solutions to tune the absorption of the bolometer to different atmospheric windows. (c) In practice, these solutions consist in adding a thin dielectric  layer (for example silicon) on the top of an existing PACS type array.}
   \end{figure}

Our bolometers use a resonant cavity to absorb submillimeter waves. It is based on a principle which is known for a long time (see [\citenum{Hadley47}]). When an absorbing material\footnote{We use a  TiN alloy  as the absorber because it is compatible with the manufacture processes  and superconductive at 300mK but absorbing for submillimeter waves.} is placed at a distance $\lambda/4$ over a reflecting layer, it is in theory possible to absorb 100\% of the energy at this particular wavelength $\lambda$. To extend the absorption profile, we use specific metallic patterns like crosses or loops (horizontal resonance, see figure \ref{absprinciple}a).

In our design, we use indium bumps to tune the cavity height to the desired wavelength we want to detect. These bumps are also used as electrical links between the silicon grid (containing the resistors) and the reflector layer where we find the first stage of signal processing.

\begin{figure}
   \begin{center}
   \begin{tabular}{c}	
    	\subfigure[]{\includegraphics[bb=210 132 623 462, height=6.75cm, clip]{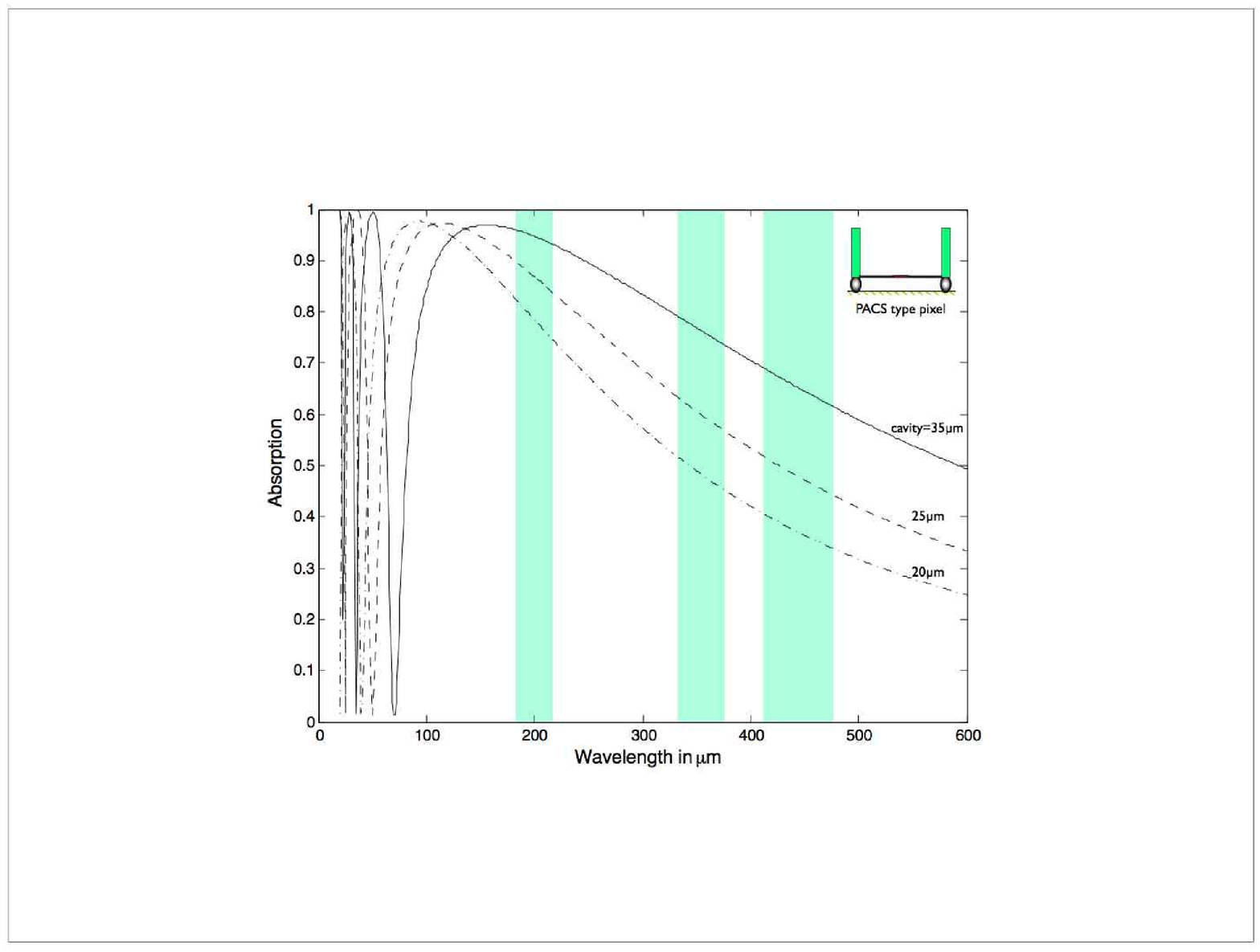}}	
  	\subfigure[]{\includegraphics[bb=210 132 623 462, height=6.75cm, clip]{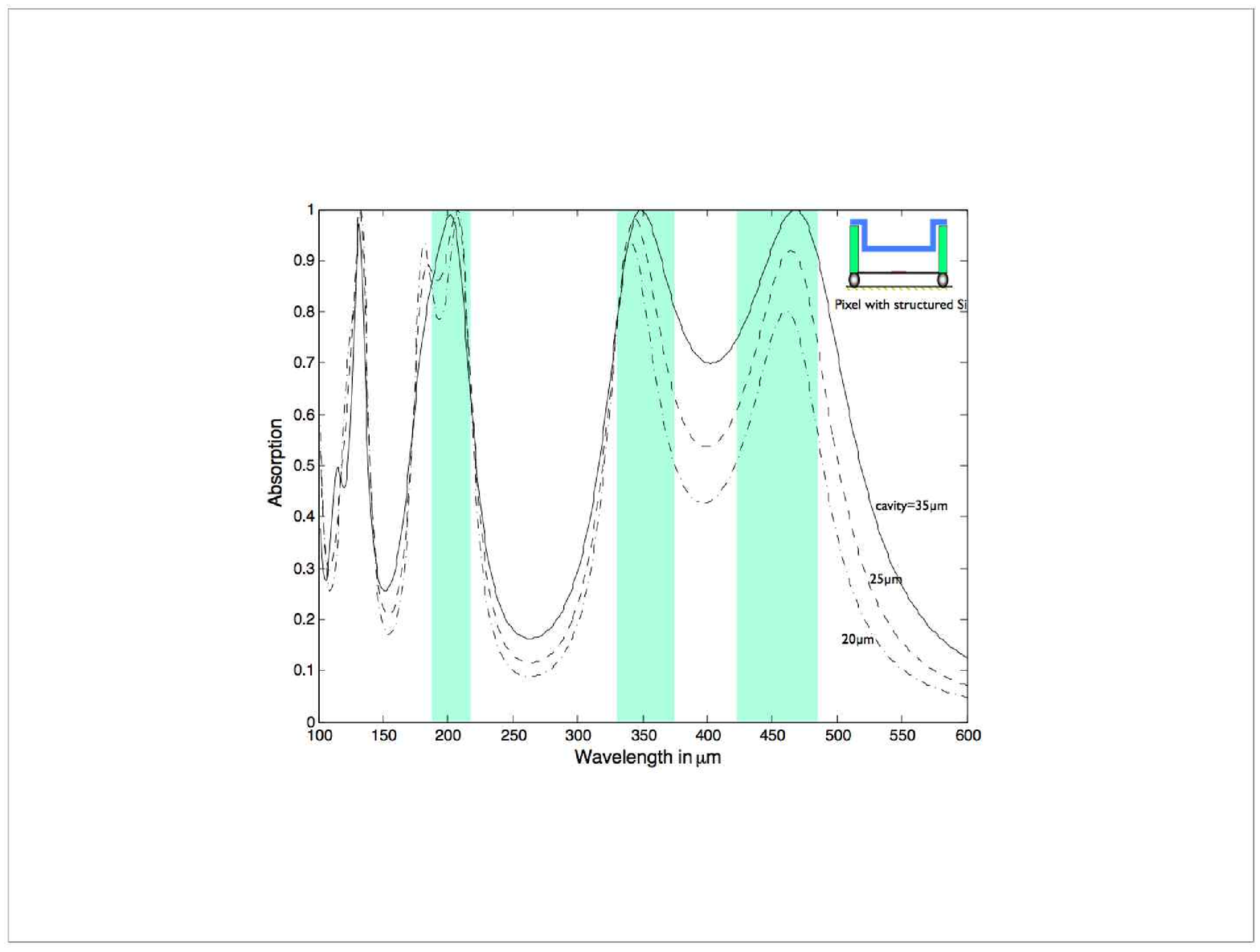}}
  \end{tabular}
   \end{center}
   \caption[example] 
   { \label{3bands} 
(a) Simulated absorption curves when using a PACS bolometer, for three cases of cavity thickness. (b) When using a strucutred silicon layer of 59 $\mu$m, a bolometer can have simultaneously a maximum absorption coefficient in the three atmospheric bands at 200, 350 and 450 $\mu$m.}
   \end{figure}

With our current manufacturing methods, there is a limitation in the size of these indium bumps, around 35 $\mu$m (PACS uses two types of cavities, at 20 and 25 $\mu$m, adapted respectively to detection at 100 $\mu$m and 180 $\mu$m). Even with the help of the metallic patterns, it is not possible to have a very good absorption in bands over 300 $\mu$m (see figure \ref{3bands}a).

In 2000, we started to study different technological possibilities to overcome this problem. One possible solution is to add a dielectric layer on the front side of the bolometers. When optimizing the thickness of this layer, it acts as an antireflective system, which reduces absorption at certain wavelengths and enhances it at others. It also has the great advantage of keeping the existing PACS type bolometers design by only adding a dielectric layer on the front side of it. We have chosen silicon as the antireflective material. The layer can be either flat or structured, in order that we can tune the second cavity to different heights (noted B in figure \ref{absprinciple}b).

To modelise the absorption, we have developed a 1D approach of this optical system, based on thin layers formalism. One way to describe the light behaviour in a stack of dielectric layers is to use an algorithm based on "characteristic matrix". Each layer in the stack is modelised by a 2$\times$2 matrix which describes how the electric field varies between the two sides of a particular dielectric layer. If we know the electric field at the entrance, it is therefore possible to predict it at the exit of the stack, by simply multiplying the matrixes. It is then possible to get access to the transmission and reflexion coefficients. 

In our case, each layer\footnote{In our model, we consider the following layers : silicon(antireflective), cavity B, absorber, silicon grid, cavity A and reflector.} is modelised by such a matrix.  We then vary free parameters like the thickness of the silicon antireflective layer and the height of the cavity B  to find solutions where the reflection coefficient of the stack would be zero, for a given optical bandwidth. The figure \ref{abscontours}a presents a diagram where contours show the absorption coefficient when the thickness of the silicon layer varies. Many solutions exist for a given wavelength, and they are chosen relatively to their technical feasibility.

\subsection{Results} 

\begin{figure}
   \begin{center}
   \begin{tabular}{c}	
    	\subfigure[]{\includegraphics[bb=181 146 613 469, height=6.5cm, clip]{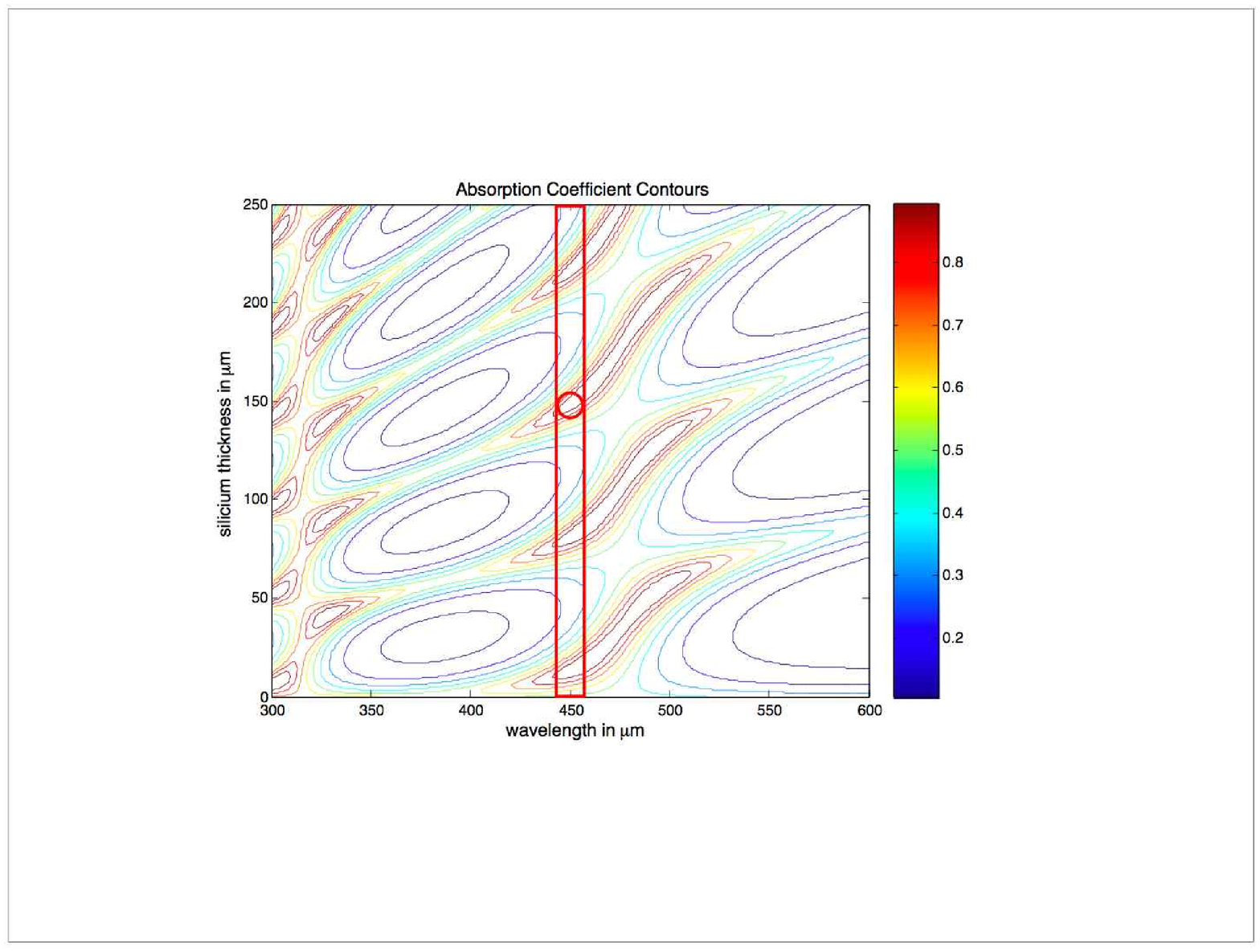}}
  	\subfigure[]{\includegraphics[bb=256 179 584 428, height=6.3cm]{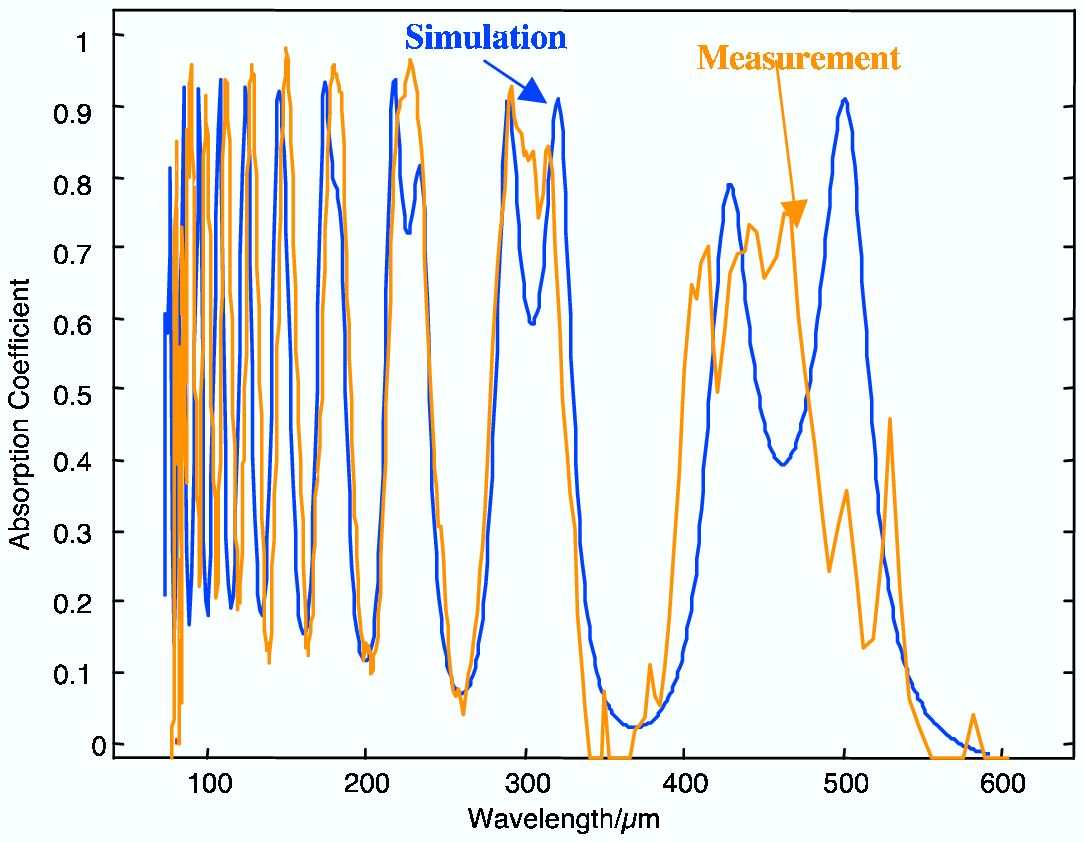}}
   \end{tabular}
   \end{center}
   \caption[example] 
   { \label{abscontours} 
(a) Diagram showing the absorption coefficient contours of a bolometer as a function of the wavelength and the thickness of the dielectric layer. The red circle shows a possible solution for a maximum absorption at 450 $\mu$m (silicon layer of 148 $\mu$m). (b) Experimental verification of the predicted absorption profile in the case of plane silicon layer of 148 $\mu$m.}
   \end{figure}

To test the performances of this concept, we have decided to use an existing PACS array (figure \ref{absprinciple}c) originally adapted to the [130 - 210] $\mu$m PACS band (cavity A of 25 $\mu$m) and which has been modified in order to get a high absorption coefficient in the 450 $\mu$m atmospheric band. In figure \ref{abscontours}a, one possible solution is to use a plane silicon layer with a thickness of 148 $\mu$m, glued on the top of the array.

Figure \ref{abscontours}b displays the expected  profile and shows indeed an absorption peak around the 450 $\mu$m central wavelength. A cold Martin-Puplett Fourier Transform Spectrometer\footnote{A Martin-Puplett Fourier Transform Spectrometer is a particular type of Michelson interferometer, in which  the beam splitter is a polarizing grid. It is well suited for studies over broad bands in submillimeter domain, because the polarization properties remain constant.} has been used to verify experimentally this absorption curve. The experimental curve in figure \ref{abscontours}b shows a very good correspondance to the expected simulated profile\footnote{Other types of models have been used to verify this kind of design, like 3D electro-magnetic modelling with HFSS, done at CEA/LETI.}, which gives a lot of credit to this new method.

Using this one dimension approach, we tried to find solutions to adapt a PACS array to the other submillimeter atmospheric bands. We found  that, when adding a 59 $\mu$m silicon layer at a distance of 175 $\mu$m above the absorbing grid, the bolometer can be simultaneously sensitive to the three bands at 200, 350 and 450 $\mu$m. This is a very interesting feature, because we can imagine  building a single focal plane instrument which could be background limited  in the three bands, using different passband optical filters. Solutions exist for adapting the arrays to the 850 $\mu$m and 1.2 mm bands, but give less efficient results in terms of absorption, due to the size limitation of the quarter-wave cavity.

To build this particular "three bands" array, the lower cost solution consists in etching a structured silicon layer which will be embedded inside each pixel\footnote{The inter-pixel silicon walls are 450 $\mu$m high and can not be reduced in size at low cost. That justifies the use of a structured antireflective silicon layer.}. Tests have been made at CEA/LETI, using RIE etching on double SOI (Silicon On Insulator) substrates and have successfully demonstrated  the feasibility of this technique.

\section{Conclusions} 

CEA is now developing bolometer arrays for ground-based applications, thanks to all  the experience gained with the development of the PACS bolometers.  A specific program, called ARTEMIS is totally dedicated to that issue (see M. Talvard et al., this volume). Simulations and laboratory work have already started and first results show that it is possible to use the PACS arrays for atmospheric bands between 200 and 450 $\mu$m with very small adjustments (thermal conductance). We show that the bolometers can be background limited in all these bands. The other interesting result concerns a new absorption system that  makes possible for a single array to be used in all the three bands. Tests on ground-based telescope have started, with such modified arrays and show promising results.

\bibliography{report}   
\bibliographystyle{spiebib}   

\end{document}